\documentclass[english,preprint]{aastex}
\usepackage[T1]{fontenc}
\usepackage[latin9]{inputenc}
\usepackage{amssymb}
\usepackage{esint}

\makeatletter
\usepackage{babel}
\makeatother

\begin{document}

\title{Probing the Turbulence Dissipation Range and Magnetic Field Strengths
in Molecular Clouds}

\author{Hua-bai Li}

\affil{Harvard-Smithsonian Center for Astrophysics, 60 Garden Street, Cambridge,
MA 02138, USA}

\email{hli@cfa.harvard.edu}

\author{Martin Houde}

\affil{Department of Physics and Astronomy, The University of Western Ontario,
London, Ontario, Canada N6A 3K7}

\email{houde@astro.uwo.ca}


\begin{abstract}
We study the turbulent velocity dispersion spectra of the coexistent
HCN and HCO$^{+}$ molecular species as a function of length scale
in the M17 star-forming molecular cloud. We show that the observed
downward shift of the ion\textquoteright{}s spectrum relative to that
of the neutral is readily explained by the existence of an ambipolar
diffusion range within which ion and neutral turbulent energies dissipate
differently. We use these observations to evaluate this decoupling
scale and show how to estimate the strength of the plane-of-the-sky
component of the embedded magnetic field in a completely novel way.
\end{abstract}

\section{Introduction}

Star formation is one of the very basic topics of astronomy. It was
proposed as early as in the 18th century, by Immanuel Kant and Simon
de Laplace independently, that stars originate from clouds of matter
contracting under their own gravitation. This point of view is supported
by modern observations, which show a high correlation between the
positions of newly formed stars and those of molecular clouds. Molecular
clouds are made up primarily of molecular hydrogen with a mean number
density of approximately $10^{2}$ cm$^{-3}$. Some of the clouds
are as big as 10 \textendash{} 100 pc with masses reaching $10^{4}$
\textendash{} $10^{6}$$M_{\odot}$, while the typical temperature
of an individual cloud is only 10 \textendash{} 50 K. Since the Jeans
mass, which is the minimum cloud mass needed to overcome thermal motion
and bring gravitational collapse, is approximately $20\, M_{\odot}$,
it follows that molecular clouds are highly gravitationally bound.
On the other hand, the star formation efficiency, defined as the ratio
of stellar mass to host cloud mass, is observed to be only a few percents
\citep{Myers 1986}. Moreover, the free fall time of a molecular cloud
(on the order of 1 Myr) is usually shorter than its age (approximately
2 \textendash{} 10 Myr). These observations imply the existence of
mechanisms that prevent the rapid collapse of molecular clouds and
explain the low efficiency with which stars are formed.

Different mechanisms have been invoked to regulate star formation.
If the potential for magnetic fields embedded in molecular clouds
to bring support against self-gravity was recognized early on \citep{Shu 1987, Mouschovias 1999},
the role played by interstellar turbulence in the star-forming process
was also put forth and is actively studied \citep{Mac Low 2004}.
The fact that such fundamental questions are still debated owes largely
to the lack of observational constraints on existing star formation
theories. For example, our knowledge of magnetic field strengths in
molecular clouds is mainly based on observations of the Zeeman effect
\citep{Crutcher 1993} and the so-called Chandrasekhar-Fermi method
\citep{Chandra 1953}, but both types of measurements are difficult
to carry out in most parts of a cloud. Similarly, turbulent energy
needs to dissipate at small scales to allow for cloud material to
collapse and form stars. But the details pertaining to the mechanisms
at the origin of turbulence dissipation, whether they are related
to viscosities or ambipolar diffusion, are still uncertain. 

In this paper, we show how the difference between the turbulent velocity
dispersion spectra of coexistent ions and neutrals can shed light
on the existence of an ambipolar diffusion range, and be used to estimate
magnetic field strengths in molecular clouds. To do so, we first use
the method put forth by \citet{Ostriker 2001} to study simulations
of turbulent molecular clouds to measure the spectral index of the
Kolmogorov-type turbulence observed in our velocity dispersion spectra.
We then argue that the observed difference between the ion and neutral
spectra is due to differences in their respective energy dissipation
ranges. Based on the measured spectral index, and the difference in
the turbulence spectra, we describe how the strength of magnetic fields
can be estimated.

\section{Turbulence in Molecular Clouds }

The observations of molecular line profiles in most parts of giant
molecular clouds (GMCs) reveal much wider line widths (on the order
of 1 km/s and more) than what is expected from thermal motions (a
fraction of 1 km/s). While there is no evidence that this motion is
due to systematic mass motion, e.g., gravitational collapse, \citet{Zuckerman 1974}
suggested that the large line widths stem from localized supersonic
turbulence. This idea is strongly supported by the discovery by \citet{Larson 1981}
that the line width follows a Kolmogorov-type spectral law, i.e.,
$\sigma\varpropto L^{0.38}$, where $\sigma$ stands for the velocity
dispersion measured from line profiles and $L$ is the linear size
of the observed clouds or cloud cores (see the Fig. 1 of \citet{Larson 1981}).
Larson\textquoteright{}s law was derived from the statistics of many
clouds/cores of different sizes, but the characteristics of the turbulence
velocity spectrum as a function of the length scale vary from cloud
to cloud. This can be seen from Larson\textquoteright{}s work, where,
for example, data pertaining to the M17 molecular cloud and cores
embedded within it imply a smaller spectral index than that (i.e.,
0.38) of the overall law mentioned above.

\begin{figure}

\plotone{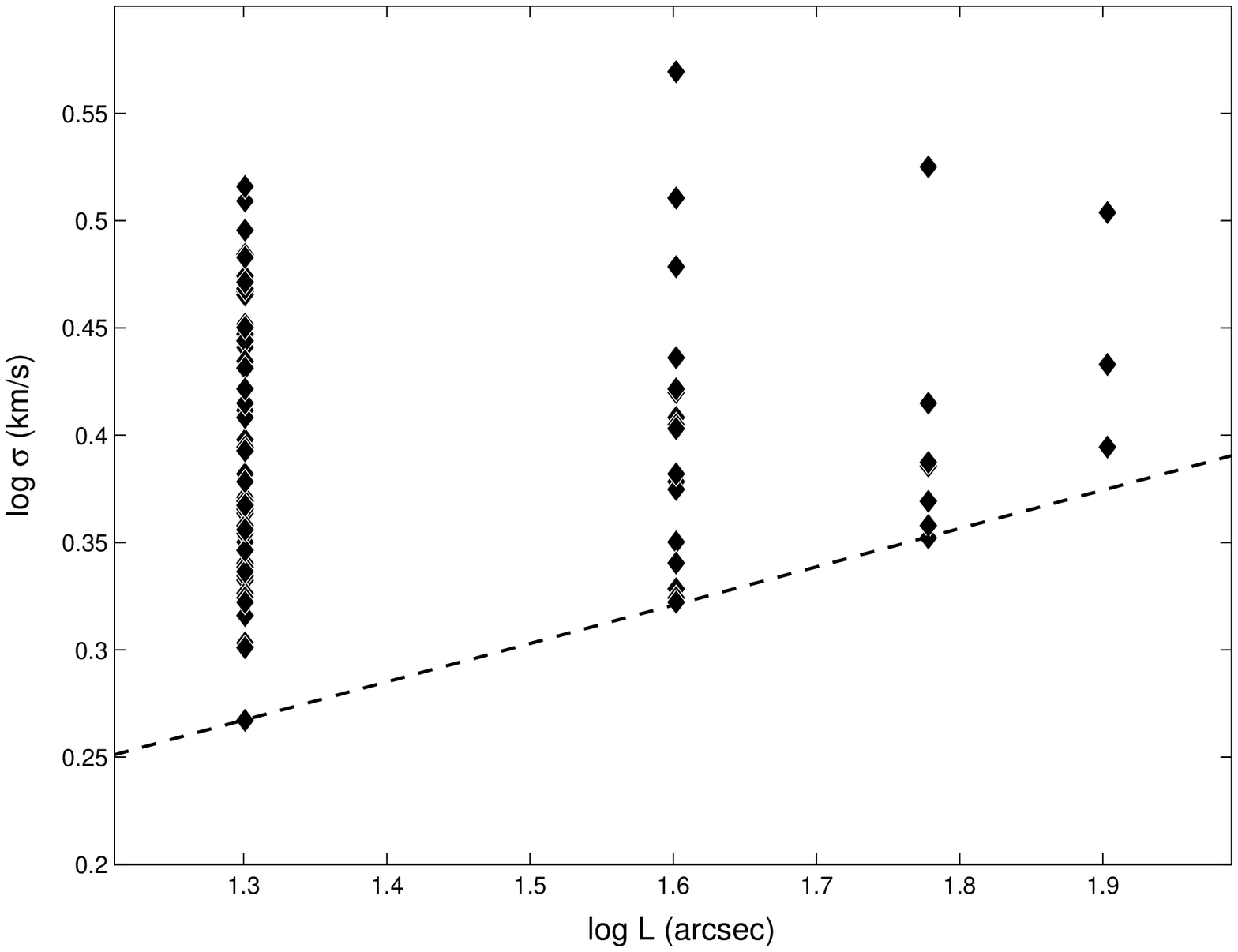}

\caption{\label{fig:HCN}The HCN velocity dispersions $\sigma$ obtained from
spectral line width measurements at sub-cloud scales $L$ (using logarithmic
scales). The lower envelope can be fitted to a Kolmogorov-type spectrum
with a spectral index of approximately 0.18, as measured from the
logarithmic slope. The observed data are at the smallest scale shown
(i.e., 20 arcseconds), while adjacent positions were combined to get
velocity dispersions at larger scales. }

\end{figure}

Several methods have been proposed to study turbulence at sub-cloud/core
scales \citep{Ossenkopf 2002}. We apply here the technique introduced
by \citet{Ostriker 2001} to our data obtained for M17. Our data consist
of the HCN $\left(J=4\rightarrow3\right)$ and HCO$^{+}$ $\left(J=4\rightarrow3\right)$
rotational transitions observed in an approximately 2 arcmin. $\times$
2 arcmin.-wide region located near the center of M17 with the Caltech
Submillimeter Observatory (CSO). These data were previously discussed
in details by \citet{Houde 2002}. Figure \ref{fig:HCN} shows a log-log
plot of the HCN velocity dispersion spectrum as a function of the
length scale. The angular resolution of the CSO\textquoteright{}s
10.4-meter telescope is approximately 20 arcseconds at the frequencies
of interest (i.e., approximately 355 GHz and 357 GHz, respectively,
for HCN and HCO$^{+}$ at this transition), and corresponds to the
smallest scale on the plot. The data at larger scales were obtained
by combining the spectra from adjacent regions to effectively simulate
beam sizes of double, triple, and quadruple the original size. In
Larson\textquoteright{}s work it was adequate to approximate the clouds
or cores as being spherical in shape, and scales along the line of
sight could, therefore, be reasonably estimated from corresponding
spatial extents on the sky. On the other hand, at sub-cloud/core scales
the sky-projected and line-of-sight scales are no longer directly
correlated. Consequently, the scatter at each scale in our data of
Figure \ref{fig:HCN} is mainly due to the range of line-of-sight
scales. But, interestingly, \citet{Ostriker 2001} showed in their
simulations of turbulent molecular clouds that the lower envelope
of such velocity dispersion distributions traces the actual turbulent
velocity spectrum very well. 

Two points are worth noting. First, the lower envelope of our HCN
data shows a spectral index of 0.18 (as measured from the logarithmic
slope), which is smaller than that observed by \citet{Larson 1981}
for his ensemble of molecular clouds. Second, in contrast to the simulations
of \citet{Ostriker 2001}, this spectral index is constant across
the dispersion spectrum. The change in spectral index at smaller scales
seen in the results of \citet{Ostriker 2001} (see their Figure 10(a))
is explained by the presence of a turbulent energy dissipation scale.
Accordingly, we interpret the constant spectral index of the HCN lower
envelope in our data as an indication that we did not reach the dissipation
scale in M17 at the spatial resolution with which our data were obtained.
That is, our data probe the inertial range.

\section{The Ambipolar Diffusion Scale }

It is usually believed that magnetic fields in the interstellar medium
are highly coupled to the gas; this is the so-called \textquotedblleft{}flux
freezing\textquotedblright{} phenomenon. This is one of the most important
consequences of the induction equation

\begin{equation}
\frac{\partial\mathbf{B}}{\partial t}=\nabla\times\left(\mathbf{v\times\mathbf{B}}\right)+\eta\nabla^{2}\mathbf{B},\label{eq:induction}\end{equation}

\noindent where $\mathbf{B}$ is the magnetic field, $\mathbf{v}$
the velocity of the (ionized) gas, and $\eta$ the magnetic diffusivity
(assumed to be constant here). The ratio of the convection term to
the diffusion term is defined as the magnetic Reynolds number

\begin{equation}
R_{\mathrm{m}}=\frac{\left|\nabla\times\left(\mathbf{v\times\mathbf{B}}\right)\right|}{\left|\eta\nabla^{2}\mathbf{B}\right|}\sim\frac{VL}{\eta},\label{eq:Rm_1}\end{equation}

\noindent where $L$ and $V$ are characteristic spatial and velocity
scales, respectively. Under conditions when $R_{\mathrm{m}}\gg1$,
the flux freezing approximation is warranted through Kelvin\textquoteright{}s
vorticity theorem \citep{Choudhuri}, as the convection term dominates
over the diffusion term in equation (\ref{eq:induction}).

When dealing with weakly ionized plasmas, however, the diffusivity
is not limited to that conveyed through $\eta$. The corresponding
analysis reveals a more general form for Ohm\textquoteright{}s law
\citep{Parker}

\begin{equation}
\mathbf{E}=\frac{1}{c}\left[-\mathbf{v}_{\mathrm{n}}\times\mathbf{B}+\frac{4\pi}{c}\left(\eta\mathbf{j}+\alpha\mathbf{j}\times\frac{\mathbf{B}}{B}+\beta\mathbf{j}_{\bot}\right)\right],\label{eq:E}\end{equation}

\noindent where $c$ is the speed of light, $\mathbf{v}_{\mathrm{n}}$
is the velocity of the neutral component of the gas, $\mathbf{j}$
and $\mathbf{j}_{\bot}$ are the current density and its component
perpendicular to $\mathbf{B}$, while $\alpha$ and $\beta$ are,
respectively, the Hall coefficient and the effective magnetic diffusivity
(also known as the Pedersen coefficient \citep{Parker}). It can be
shown that for typical conditions encountered in GMCs $\beta\gg\alpha\gg\eta$,
and the above relation for the electric field thus simplifies to \citep{Biskamp}

\begin{equation}
\mathbf{E}=-\frac{1}{c}\left(\mathbf{v}_{\mathrm{n}}\times\mathbf{B}-\frac{4\pi}{c}\beta\mathbf{j}_{\bot}\right).\label{eq:E_eff}\end{equation}

The effective magnetic diffusivity is given by

\begin{equation}
\beta=\frac{B^{2}}{4\pi n_{\mathrm{i}}\mu\nu_{\mathrm{i}}},\label{eq:beta}\end{equation}

\noindent where $n_{\mathrm{i}}$ is the ion density, while $\nu_{\mathrm{i}}$
and $\mu$ are the collision rate of an ion with the neutrals and
the mean reduced mass characterizing such collisions, respectively.
Although the insertion of equation (\ref{eq:E_eff}) into Faraday\textquoteright{}s
law of induction does not yield a relation analogous to equation (\ref{eq:induction})
for the induction equation \citep{Brandenburg 1994}, it is nonetheless
clear that the flux freezing approximation will cease to be adequate
when the effective magnetic diffusivity term dominates over the convective
term in equation (\ref{eq:E_eff}). Consequently, we can define a
new effective magnetic Reynolds number with 

\begin{eqnarray}
R_{\mathrm{m}} & = & \frac{c\left|\mathbf{v_{\mathrm{n}}\times\mathbf{B}}\right|}{4\pi\left|\beta\mathbf{j}_{\bot}\right|}=\frac{\left|\mathbf{v_{\mathrm{n}}\times\mathbf{B}}\right|}{\left|\beta\left(\nabla\times\mathbf{B}\right)_{\bot}\right|}\nonumber \\
 & \sim & \frac{V_{\mathrm{n}}L}{\beta}\nonumber \\
 & \sim & \frac{4\pi n_{\mathrm{i}}\mu\nu_{\mathrm{i}}V_{\mathrm{n}}L}{B^{2}},\label{eq:Rm2}\end{eqnarray}

\noindent where Ampère\textquoteright{}s law was substituted for the
current density. Based on equation (\ref{eq:Rm2}), it is clear that
although $R_{\mathrm{m}}\gg1$ for scales usually probed with single-dish
observations (i.e., on the order of sizes subtended by the telescope
beam), this relation does not necessarily hold true for all the turbulent
eddies existing in any given region of a molecular cloud. This is
because situations involving small eddies (in the neutral gas) of
size $L$ evolving in a region embedded with a (strong) magnetic field
will lead to lower effective magnetic Reynolds numbers. In fact, it
is expected from the loss in the relative importance of the convective
term in equation (\ref{eq:E_eff}) that the magnetic field, and therefore
the ions, will decouple more easily from neutral eddies at such scales,
when the effective Reynolds number is on the order of unity or less.
This decoupling, or ambipolar diffusion \citep{Mestel 1956}, will
be responsible for friction forces between the ion and neutral components
of the gas, and will thus bring about the dissipation of turbulent
energy \citep{Zweibel 1983}. It is also reasonable to expect that
the existence of ambipolar diffusion will be the cause for possible
differences between the velocity dispersion spectra of the ions and
the neutrals. 

We also note that for typical conditions in the interstellar medium
(e.g., $B\sim10\,\mu\mathrm{G}$, $n_{\mathrm{n}}\sim10^{3}$ cm$^{-3}$,
an ionized fraction of approximately $10^{-6}$, and with $\nu_{\mathrm{i}}\simeq1.5\times10^{-9}n_{\mathrm{n}}$
s$^{-1}$ for ion/neutral collisions \citep{Nakano 1984}) equation
(\ref{eq:beta}) yields an effective magnetic diffusivity $\beta$
of the order of $10^{21}$ cm$^{2}$s$^{-1}$, which is much larger
than the hydrodynamic viscosity (which is of the order of $10^{16}$
cm$^{2}$s$^{-1}$). The situation also holds for denser parts of
molecular clouds such as the ones probed with our data, where both
density and magnetic field strength can increase by as much as two
or three orders of magnitude. Given this fact, will ambipolar diffusion
set an energy dissipation scale in the turbulent energy spectrum that
is larger than the scale due to viscosity? This is still an open question,
both on observational and theoretical/simulation grounds. For supersonic
magnetohydrodynamic turbulence, the fast waves and Alfvén waves are
strongly damped at smaller scales when $R_{\mathrm{m}}<1$ \citep{McKee 1993, Balsara 1996}.
It is, therefore, reasonable to expect steeper energy spectra at ambipolar
diffusion scales. But since ions and neutrals are decoupled from each
other, their respective energy (and velocity dispersion) spectra must
have different slopes at these scales. Which one should be steeper?
Here we show that observing the turbulent velocity spectra of coexistent
ion and neutral molecular species, even at spatial resolution much
lower than ambipolar diffusion scales, can shade some light on these
questions.

\section{Comparison of HCN and HCO$^{+}$ in M17}

Figure \ref{fig:HCN_HCO+} shows in part the same data as in Figure
\ref{fig:HCN}, with the exceptions that the vertical axis is now
for the square of the velocity dispersion, we used linear scales for
both the horizontal and vertical scales, and more importantly data
for the other molecular species, HCO$^{+}$, have been added. The
use of linear scales for the axes in this figure was chosen to better
show the difference between the two data sets as a function of the
length scale, as this will become important for the discussion below.
As can be seen, the two molecular species show very similar spectral
indices for their respective velocity dispersion spectra (i.e., 0.18,
as calculated with the lower envelopes), which is expected at this
scale because ions are trapped by magnetic fields, which are in turn
well-coupled with neutrals due to flux freezing. But note that the
lower envelope of HCO$^{+}$ is downshifted from that of HCN. We now
turn our attention to identifying the cause for this nearly constant
shift of the ion velocity spectrum over all the scales shown in the
part of the inertial range probed by our observations.

HCN and HCO$^{+}$ have been shown to be coexistent in molecular clouds
and to have systematic line width differences \citep{Houde 2000a, Houde 2000b, Houde 2002, Lai 2003, Houde 2004}.
Due in part to its smaller critical density HCO$^{+}$ is generally
somewhat more optically thick and will accordingly usually exhibit
slightly more extended spatial distributions than HCN \citep{Houde 2000a, Houde 2000b}.
Both of these properties would suggest that HCO$^{+}$ should have
wider line widths than HCN, yet the fact is that its spectral line
profiles are consistently narrower in turbulent molecular clouds (such
as M17). \citet{Houde 2000a} put forth a model, where magnetic fields
affect the dynamics of molecular ions through the cyclotron interaction,
that can account for the narrowing of their spectral line profiles,
which was in turn interpreted as a signature of ambipolar diffusion
\citep{Houde 2004}. 

Here we further argue that the aforementioned observed difference
between the two turbulent velocity spectra (as a function of length
scale) is also the result of ambipolar diffusion. Even though ambipolar
diffusion happens at much smaller scales than that at which our data
was obtained, it is still discernible in our observations because
the velocity dispersions we measure stem from the contribution of
all turbulent eddies with scales approximately equal to, or smaller
than, the different beam sizes. To explain the systematically downshifted
velocity dispersion spectrum of the ions, it is necessary that 

\begin{enumerate}
\item Ambipolar diffusion does not set the cut-off range for both the ion
and neutral energy spectra.
\item The ion spectral slope must be steeper than that of the neutrals\textquoteright{}
at ambipolar diffusion scales, as illustrated in Figure \ref{fig:model}
(more below). 
\end{enumerate}
As mentioned before, a steep drop in the ion velocity dispersion spectrum
at ambipolar diffusion scales can be expected due to the fact that
most of families of magnetohydrodynamic waves are strongly damped
when $R_{\mathrm{m}}<1$. Neutrals, while still partaking in the turbulent
energy cascade of the inertial range, should also exhibit a steeper
turbulent spectrum than in their inertial range because of friction
forces between ions and neutrals. But since the low ionization level
will limit the amount of friction, neutrals will still have their
energy cut-off set by hydrodynamic viscosity. The square of the observed
velocity dispersion at a particular scale is proportional to the integral
of the energy spectrum below that scale. Therefore, at all scales
larger than those covered by the ambipolar diffusion range the difference
between the squares of the neutral and ion velocity dispersions will
be proportional to the difference of the corresponding integrals between
the scale where ambipolar diffusion sets in and the hydrodynamic dissipation
scale (see below). This is conveyed through the shaded region in Figure
\ref{fig:model}.

\section{Estimation of the Magnetic Field Strength }

The difference between the ion and neutral velocity dispersion spectra
can be used to estimate the strength of the magnetic field. To see
how this can be achieved, let us consider the wave number $k^{\prime}=2\pi/L^{\prime}$
corresponding to turbulent eddies of characteristic size $L^{\prime}$
such that $R_{\mathrm{m}}=V_{\mathrm{n}}^{\prime}L^{\prime}/\beta=1$,
with $V_{\mathrm{n}}^{\prime}$ the characteristic velocity of these
(neutral) eddies. As discussed earlier, this is the scale at which
ambipolar diffusion sets in. For a wave number $K<k^{\prime}$ (i.e.,
of spatial scale larger than $L^{\prime}$), the difference between
the squares of the neutral and ion velocity dispersions can be written
as

\begin{equation}
\Delta\sigma_{K}^{2}=\int_{K}^{\infty}\left[F_{\mathrm{n}}\left(k\right)-F_{\mathrm{i}}\left(k\right)\right]dk,\label{eq:DsK}\end{equation}

\noindent where $F_{\mathrm{n}}\left(k\right)$ and $F_{\mathrm{i}}\left(k\right)$
are (twice) the turbulent energy spectra density (i.e., per unit mass
and wave number) for the neutral and ion species, respectively. But
because of the good coupling of the ions and neutral through flux
freezing in the inertial range we expect that 

\begin{equation}
\begin{array}{cc}
F_{\mathrm{n}}\left(k\right)=F_{\mathrm{i}}\left(k\right), & \mathrm{for}\, k<k^{\prime}\end{array}.\label{eq:Fn=Fi}\end{equation}

If we further assume that $F_{\mathrm{i}}\left(k\right)$ decreases
steeply beyond $k^{\prime}$, then for any wave number $K<k^{\prime}$
the difference of the square of the velocity dispersions (eq. {[}\ref{eq:DsK}])
is only a function of $k^{\prime}$ with

\begin{equation}
\begin{array}{cc}
\Delta\sigma_{K}^{2}\simeq\Delta\sigma_{k^{\prime}}^{2}\equiv\int_{k^{\prime}}^{\infty}F_{\mathrm{n}}\left(k\right)dk, & \mathrm{for}\, K<k^{\prime}\end{array}.\label{eq:Dsk'}\end{equation}

This result is consistent with our observations, since, as was mentioned
earlier, our data show such a nearly constant difference between the
neutral and ion spectra. More precisely, the measured differences
$\Delta\sigma_{K}^{2}$ from the velocity dispersion spectra of Figure
\ref{fig:HCN_HCO+} are 0.57, 0.65, 0.57, and 1.22 km$^{2}$/s$^{2}$
with increasing beam sizes. The discrepancy at the largest spatial
scale is easily accounted for by the small amount of statistics available,
which prevented us from finding the true lower envelope for both velocity
dispersion spectra.

An important consequence that follows from this analysis is that the
determination of the decoupling scale $L^{\prime}$ and the neutral
velocity dispersion $V_{\mathrm{n}}^{\prime}$ at $L^{\prime}$ would
allow for the evaluation of the magnetic field strength through equation
(\ref{eq:Rm2}) by setting $R_{\mathrm{m}}=1$. More precisely, we
find that

\begin{equation}
B^{2}=4\pi n_{\mathrm{i}}\mu\nu_{\mathrm{i}}V_{\mathrm{n}}^{\prime}L^{\prime},\label{eq:B^2}\end{equation}

\noindent which can be conveniently transformed using typical parameters
for GMCs. For example, if we use mean neutral and ion masses of, respectively,
2.3 and 29 times that of the hydrogen atom, then

\begin{equation}
B=\left(\frac{L^{\prime}}{0.5\,\mathrm{mpc}}\right)^{1/2}\left(\frac{V_{\mathrm{n}}^{\prime}}{1\,\mathrm{km\, s^{-1}}}\right)^{1/2}\left(\frac{n_{\mathrm{n}}}{10^{6}\,\mathrm{cm^{-3}}}\right)\left(\frac{\chi_{\mathrm{i}}}{10^{-7}}\right)^{1/2}\,\mathrm{mG,}\label{eq:B}\end{equation}

\noindent where $\chi_{\mathrm{i}}$ is the ionization fraction.

It is important to note that since observations can only reveal the
velocity dispersion along the line-of-sight and that the effective
magnetic diffusivity term in equation (\ref{eq:E_eff}) is in a direction
perpendicular to the magnetic field, the strength of the field estimated
with equations (\ref{eq:B^2}) or (\ref{eq:B}) is only valid for
its plane-of-the-sky component. 

We now proceed and evaluate $L^{\prime}$ and $V_{\mathrm{n}}^{\prime}$
using our data. First, we note that the velocity dispersion due to
an eddy of scale size $k=2\pi/L$ is given by

\begin{equation}
V_{\mathrm{n}}^{2}\left(k\right)=F_{\mathrm{n}}\left(k\right)\Delta k.\label{eq:Vn^2}\end{equation}

The energy density $F_{\mathrm{n}}\left(k\right)$ can readily be
evaluated from our data, since we know that 

\begin{eqnarray}
\sigma_{\mathrm{n}}^{2}\left(k\right) & = & \int_{k}^{\infty}F_{\mathrm{n}}\left(\kappa\right)d\kappa\nonumber \\
 & \simeq & b\left(\frac{2\pi}{k}\right)^{n},\label{eq:sn^2}\end{eqnarray}

\noindent where $\sigma_{\mathrm{n}}\left(k\right)$ is the neutral
velocity dispersion, while $b$ and $n$ are the different fit parameters
to the neutral data (see Figure \ref{fig:HCN_HCO+}). Using the derivative
relative to $k$ on this equation and the spectral spread corresponding
to a Gaussian beam of full-width-half-magnitude $L$ to estimate $\Delta k$
($=\sqrt{8\ln\left(2\right)}/L$ ), we get from equation (\ref{eq:Vn^2})
that 

\begin{equation}
V_{\mathrm{n}}^{2}\left(L\right)\simeq0.37bnL^{n}.\label{eq:Vn^2(L)}\end{equation}

Moreover, if, as assumed, there is a significant drop of the ion spectrum
in its dissipation range (as shown in Figure \ref{fig:model}), it
then follows that the ion velocity dispersion $\sigma_{\mathrm{i}}\left(k^{\prime}\right)$
observed at $L^{\prime}$ (or $k^{\prime}=2\pi/L^{\prime}$) will
be approximately equal to $V_{\mathrm{n}}^{\prime}$. We, therefore,
write

\begin{equation}
V_{\mathrm{n}}^{\prime2}\simeq a+bL^{\prime n},\label{eq:Vn'^2}\end{equation}

\noindent where it should be clear that $-a$ corresponds to $\Delta\sigma_{k^{\prime}}^{2}$,
as defined in equation (\ref{eq:Dsk'}) and measured from our data
(see Figure \ref{fig:HCN_HCO+}). Equating equations (\ref{eq:Vn^2(L)})
(at $L^{\prime}$) and (\ref{eq:Vn'^2}) gives

\begin{equation}
L^{\prime n}=\frac{-a}{b\left(1-0.37n\right)}.\label{eq:L'^n}\end{equation}

For M17, the fits to our data yield $a\simeq-0.59\,\mathrm{km^{2}s^{-2}}$,
$b\simeq1.17\,\mathrm{km^{2}s^{-2}arcsecond}^{-n}$, and $n\simeq0.36$.
Equation (\ref{eq:L'^n}) then gives $L^{\prime}\simeq0.22^{\prime\prime}$
or 1.8 mpc at the distance of M17 (i.e., approximately 1.7 kpc), and
$V_{\mathrm{n}}^{\prime}\simeq0.30$ km/s. 

A precise determination of the magnetic field strength in M17 through
equation (\ref{eq:B}) would require an equally precise knowledge
of the gas density $n_{\mathrm{n}}$ and the ionization fraction $\chi_{\mathrm{i}}$,
which we do not possess at this time. We can, however, use reasonable
values for these parameters and provide an estimate of the magnetic
field strength to allow for comparison with previous measurements
obtained with other techniques. Therefore, choosing $n_{\mathrm{n}}=10^{6}\,\mathrm{cm^{-3}}$,
which should be appropriate for HCN and HCO$^{+}$ at the $J=4\rightarrow3$
transition, and $\chi_{\mathrm{i}}=10^{-7}$ for the ionization fraction
we calculate a plane-of-the sky magnetic field strength of $B\approx1.0$
mG from equation (\ref{eq:B}). Although this value compares well
with earlier H{\small{I}} and OH Zeeman measurements \citep{Brogan 2001}
that yielded line-of-sight magnetic field strengths as high as 0.75
mG in this molecular cloud, we should keep in mind that our estimate
is probably uncertain to within a factor of a few or even as much
as an order of magnitude. We also note that the details of our calculations
rest in part on the assumption that the ion velocity dispersion spectrum
drops steeply in the ambipolar diffusion range. In the event that
this assumption is not realized, the characteristic velocity $V_{\mathrm{n}}^{\prime}$,
and the magnetic field strength, would be overestimated through equation
(\ref{eq:Vn'^2}). 

It is important to note that current (e.g., SMA, CARMA) or future
(e.g., ALMA) interferometers could provide high enough spatial resolutions
to allow for the direct determination of the ambipolar diffusion scale
$L^{\prime}$ and the characteristic velocity $V_{\mathrm{n}}^{\prime}$
of an eddy at that scale directly from the velocity dispersion spectra.
The determination of $L^{\prime}$ would be straightforward from the
velocity dispersion spectrum of the ion, as it would be identified
with the scale at which a significant drop in velocity occurs (i.e.,
the ion cut-off point) or where the neutral and ion spectra depart
from one another (see Figure \ref{fig:model}).

\section{Summary}

\begin{figure}
\plotone{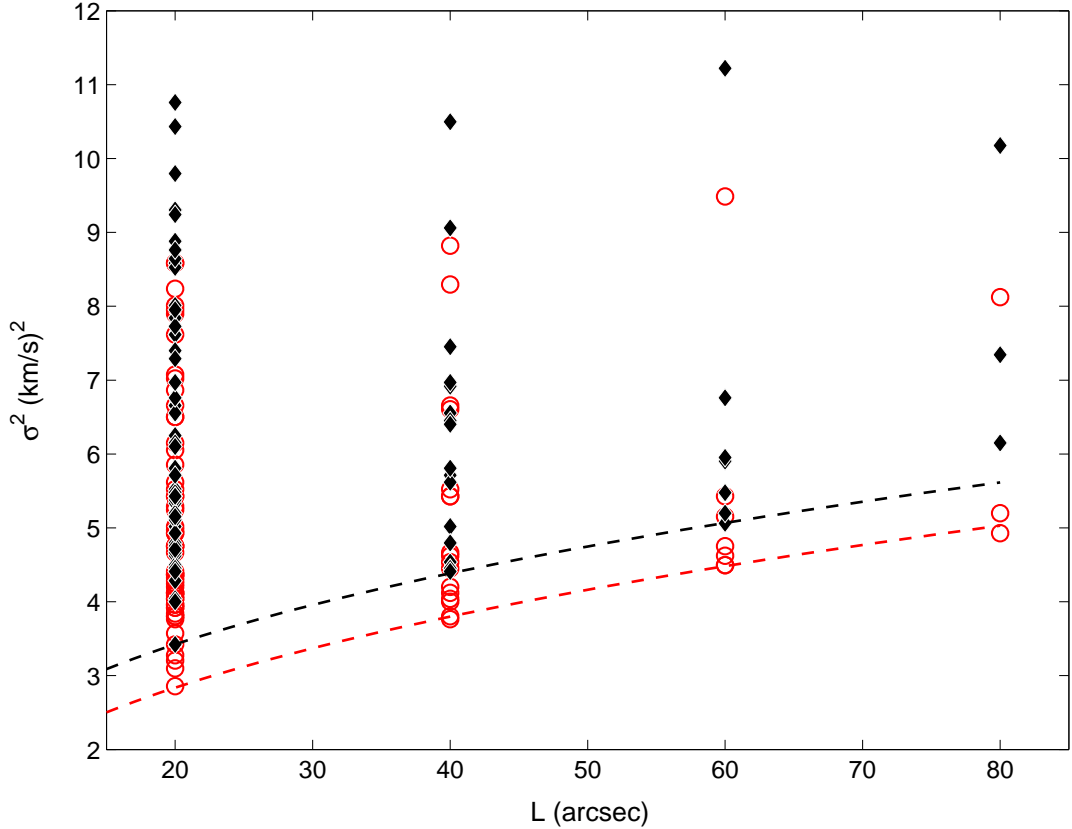}

\caption{\label{fig:HCN_HCO+}Similar to Figure \ref{fig:HCN}, with the exceptions
that the vertical axis is now for the square of the velocity dispersion,
we used linear scales for both the horizontal and vertical scales,
and the data for the other molecular species, HCO$^{+}$, have been
added (in red, HCN is in black). The HCN data are fitted for a Kolmogorov-type
law $\sigma_{\mathrm{n}}^{2}=bL^{n}$ to their lower envelope. The
relative downshift of the HCO$^{+}$ velocity dispersion spectrum
is made explicit with the corresponding $\sigma_{\mathrm{i}}^{2}=a+bL^{n}$
fit. The different parameters are $a\simeq-0.59\,\mathrm{km^{2}s^{-2}}$,
$b\simeq1.17\,\mathrm{km^{2}s^{-2}arcsecond}^{-n}$, and $n\simeq0.36$.
The measured differences between the two velocity dispersion spectra
are 0.57, 0.65, 0.57, and 1.22 km$^{-2}$s$^{-2}$ with increasing
beam sizes. The lack of agreement at the largest spatial scale is
easily accounted for by the small amount of statistics available,
which prevented us from finding the true lower envelope at that scale
for both spectra. The uncertainties on the measured velocity dispersions
for the lower envelopes are negligible ($<0.01$ km/s) and are thus
not shown.}

\end{figure}

\begin{figure}

\plotone{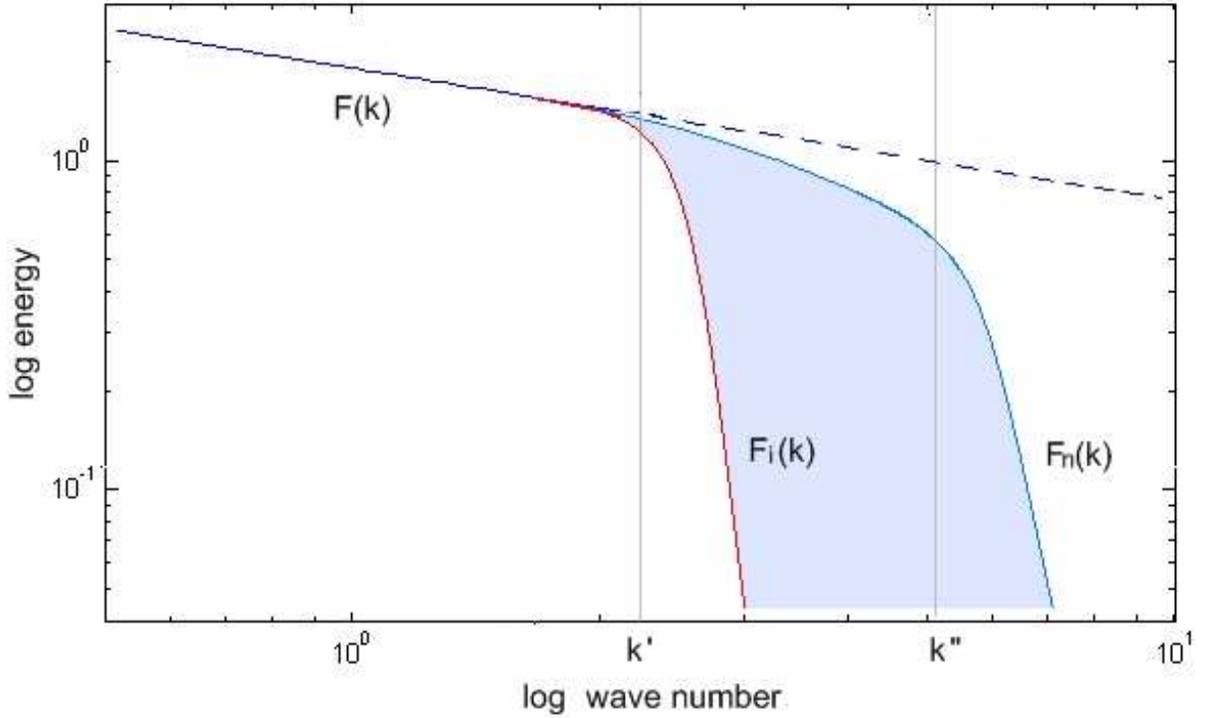}

\caption{\label{fig:model}An illustration of our explanation for the difference
between the ion and neutral velocity spectra of Figure \ref{fig:HCN_HCO+}
is shown. This is a log-log plot of the turbulent energy versus wave
number (inverse of the spatial scale) in arbitrary units. Ambipolar
diffusion happens at wave numbers larger than $k^{\prime}$, which
is in turn smaller than $k^{\prime\prime}$, where the hydrodynamic
viscosity sets in for the neutral spectrum $F_{\mathrm{n}}\left(k\right)$
($F_{\mathrm{i}}\left(k\right)$ is for the ion spectrum). The energy
spectrum in the inertial range, $F\left(k\right)$, is common to both
species, while ions and neutrals have different spectra at ambipolar
diffusion scales. The square of the velocity measured at a particular
wave number $K$ is proportional to the integral of the energy spectrum
over all wave numbers greater than $K$. The observed difference between
the two velocity dispersion spectra is proportional to the shaded
area. }

\end{figure}

Our analysis of turbulent velocity dispersion spectra from coexistent
ion and neutral molecular species can be summarized as follows: 

\begin{enumerate}
\item We showed that by combining observational spectral line profile data
to different scales, in the manner of \citet{Ostriker 2001}, it is
possible to probe turbulent velocity dispersion spectra at sub-cloud/core
scales.
\item For coexistent ions and neutrals in M17, the turbulent velocity dispersion
spectra share the same slope in the inertial range, as expected from
the flux freezing approximation, but the ion spectrum is downshifted
nearly uniformly at all scales relative to that of the neutrals, as
shown in Figure \ref{fig:HCN_HCO+}. This can be explained by the
occurrence of ambipolar diffusion at smaller scales where $R_{\mathrm{m}}<1$.
The downshifted velocity spectrum of the ions implies that they have
a steeper turbulent energy spectrum than the neutrals at ambipolar
diffusion scales. It is, therefore, apparent that ambipolar diffusion
does not set the energy cut-off range for both ions and neutrals.
\item Since ambipolar diffusion happens in the presence of magnetic fields,
the difference between the velocity dispersion spectra of ions and
neutrals can be used to estimate magnetic field strengths. For M17,
we estimate the ambipolar diffusion scale to approximately 1.8 mpc,
and the plane-of-the-sky magnetic field strength to approximately
1 mG. Because we do not precisely know the gas density and the ionization
fraction in M17 this estimate of the magnetic field strength is probably
uncertain to within a factor of a few or even as much as an order
of magnitude.
\end{enumerate}
We note that using optically thin molecular species, such as the H$^{13}$CN
and H$^{13}$CO$^{+}$ isotopologues, could be advantageous for this
type of study; their systematic line with difference is also reported
by \citet{Houde 2000b} and \citet{Lai 2003}. Their advantage would
reside in the fact that their corresponding line profiles would not
be saturated, unlike the species used in this work in some parts of
the molecular cloud under consideration, and the line widths measured
would reflect more faithfully the true velocity dispersions of the
ion and neutral gas components along the line-of-sight. Moreover,
H$^{13}$CN and H$^{13}$CO$^{+}$ would allow us to probe magnetic
fields at greater depth in star-forming regions, where they are likely
to be stronger \citep{Crutcher 1999, Basu 2000}.

\acknowledgements{The authors thank S. Basu, R. H. Hildebrand, E. Keto, P. C. Myers,
and E. C. Ostriker for insightful discussions and comments. H. Li\textquoteright{}s
research is funded through a postdoctoral fellowship from the Smithsonian
Astrophysical Observatory. M. H.'s research is funded through the
NSERC Discovery Grant, Canada Research Chair, Canada Foundation for
Innovation, Ontario Innovation Trust, and Western's Academic Development
Fund programs. The Caltech Submillimeter Observatory is funded through
the NSF grant AST 05-40882 to the California Institute of Technology. }

\end{document}